\documentclass[11pt,english,epsf]{scrartcl}
\usepackage{lmodern}
\usepackage[T1]{fontenc}
\usepackage[latin9]{inputenc}
\usepackage{geometry}
\usepackage{amsmath}
\usepackage{amssymb}

\newtheorem{theorem}{Theorem}
\newtheorem{lemma}{Lemma}
\newcommand {\dfn} {\stackrel{\Delta} {=}}
\newcommand {\exe} {\stackrel{\cdot} {=}}
\newcommand {\lexe} {\stackrel{\cdot} {\le}}

\newcommand {\bx} {\mbox{\boldmath $x$}}
\newcommand {\by} {\mbox{\boldmath $y$}}

\newcommand {\bD} {\mbox{\boldmath $D$}}
\newcommand {\bE} {\mbox{\boldmath $E$}}

\newcommand {\bR} {\mbox{\boldmath $R$}}

\newcommand {\bX} {\mbox{\boldmath $X$}}
\newcommand {\bY} {\mbox{\boldmath $Y$}}

\newcommand{\calA}{{\cal A}}

\newcommand{\calC}{{\cal C}}
\newcommand{\calD}{{\cal D}}

\newcommand{\calI}{{\cal I}}

\newcommand{\calQ}{{\cal Q}}
\newcommand{\hQ}{{\hat{Q}}}
\newcommand{\tQ}{{\tilde{Q}}}
\newcommand{\calR}{{\cal R}}

\newcommand{\calT}{{\cal T}}

\newcommand{\calX}{{\cal X}}
\newcommand{\calY}{{\cal Y}}

\allowdisplaybreaks

\begin{document}
\thispagestyle{empty}
\title{Codeword or Noise? Exact Random Coding Exponents for Slotted
Asynchronism\thanks{
This research was supported by the Israel Science Foundation (ISF),
grant no.\ 412/12.}}
\author{Neri Merhav}
\maketitle

\begin{center}
Department of Electrical Engineering \\
Technion - Israel Institute of Technology \\
Technion City, Haifa 32000, ISRAEL \\
E--mail: {\tt merhav@ee.technion.ac.il}\\
\end{center}
\vspace{1.5\baselineskip}
\setlength{\baselineskip}{1.5\baselineskip}

\begin{center}
{\bf Abstract}
\end{center}
\setlength{\baselineskip}{0.5\baselineskip}
We consider the problem of slotted asynchronous coded communication, where
in each time frame (slot), the transmitter is either silent or transmits a codeword 
from a given (randomly selected) codebook. The task of the decoder is to
decide whether transmission has taken place, and if so, to decode the
message. We derive the optimum detection/decoding rule in the sense of the
best trade-off among the probabilities of decoding error, false alarm, and
misdetection. For this detection/decoding rule, we then derive single--letter
characterizations of the exact exponential rates of these three probabilities
for the average code in the ensemble.\\

\vspace{0.2cm}


\setlength{\baselineskip}{2\baselineskip}

\newpage
\section{Introduction}

The problem of synchronization has been a long--standing, important issue in
communication throughout several decades (see, e.g., \cite{Barker53},
\cite{GDRTS63}, \cite{Franks80}, \cite{Massey72}, \cite{Scholtz80},
\cite{TCW08}, \cite{TKW06}, \cite{Wang10}, \cite{WCCW11} and references
therein, for a non--exhaustive sample of earlier works). 

The general problem
setting under consideration allows the transmitter to send messages only part
of the time, and to be `silent' (non--transmitting) when it has no messages
ready to be conveyed. The receiver then has to be able to reliably detect the
existence of the message, locate its starting time 
instant, and decode it. The traditional approach has been to separate the
problems of synchronization and coding/decoding, where in the former, a
special pattern of symbols (synchronization word) is used to mark the
beginning of a message transmission. This transmission of a
synchronization word is, however, is an undesired overhead.

Following \cite{Wang10} and \cite{WCCW11}, in this work, we treat the
synchronization and coding jointly and we adopt the simplified model of {\it
slotted} communication. According to this model, a transmission can
start only at time instants that are integer multiples of the slot length, which
is also the block length. Thus,
in each slot (or
block), the transmitter is either entirely silent, or it transmits a codeword
corresponding to one of $M$ possible messages. In the silent mode, it is
assumed that the transmitter repetitively feeds 
the channel by a special channel input symbol denoted by `$0$' (indeed, in the case
of a continuous input alphabet, it is natural to assign a zero input signal), and then
the channel output vector is thought of as ``pure noise.'' The decoder in turn
has to decide whether a message has been sent or the received channel
output vector is pure noise. In case it decides in favor of the former, it
then has to decode the message. 

In \cite{Wang10} and \cite{WCCW11}, three
figures of merit were defined in order to judge performance: (i) the probability
of {\it false alarm} (FA) -- i.e., deciding that a message has been sent when actually, the
transmitter was silent and the
channel output was pure noise, (ii) the probability of {\it misdetection} (MD)
-- that is, deciding
that the transmitter was silent when it actually transmitted some message, and
(iii) the probability of {\it decoding error} (DE) -- 
namely, not deciding on the correct message sent.
Wang \cite{Wang10} and Wang {\it et al.}
\cite{WCCW11} have posed the problem of
characterizing the best achievable region of the
error exponents associated with these three probabilities for a given discrete
memoryless channel (DMC).
It was stated in \cite{WCCW11} that this general problem is open, and so, the focus
both in \cite{Wang10} and \cite{WCCW11} 
was directed to the narrower problem of trading off the FA exponent and the MD
exponent when the DE exponent constraint is completely relaxed, that is, there
is no demand on exponential decay rate of the DE probability. Upper and lower
bounds on the maximum achievable FA exponent for a given MD exponent were
derived in these works. In the extreme case where the MD exponent constraint is omitted
(set to zero), these bounds coincide, and so, the characterization of the best
achievable MD exponent is exact.

In this paper, we adopt the same problem setting of slotted asynchronous
communication as in \cite{Wang10} and \cite{WCCW11}.
We first derive, for a given code, the optimum detection--decoding rule that minimizes the DE
probability subject to given constraints on the FA and the MD probabilities. This
detection--decoding rule turns out to be completely different from the one
in the achievability parts of \cite{Wang10} and \cite{WCCW11}. In particular,
denoting the codewords by $\{\bx_m\}$, the channel output vector by $\by$
(all of length $n$),
and the channel conditional probability by $W(\by|\bx_m)$, then according to
this rule, a transmission is
detected iff 
\begin{equation}
e^{n\alpha}\sum_{m=1}^MW(\by|\bx_m)+\max_{1\le m\le M}W(\by|\bx_m)\ge
e^{n\beta}W(\by|0^n)
\end{equation}
where $\alpha$ and $\beta$ are chosen to meet the MD and FA constraints.
Of course, whenever the received $\by$ passes this test, the maximum
likelihood (ML) decoder is
applied, assuming that all messages are equiprobable {\it a-priori}. The
performance of this optimum detector/decoder is analyzed under the random coding
regime of fixed composition codes, 
and the achievable trade-off between the three error exponents is given in
full generality, that is, not merely in the margin where at least one of the
exponents vanishes. It should be pointed out that our analysis technique, which is based on type class
enumeration (see, e.g., \cite{Merhav09}, \cite{SBM11} and references therein), 
provides the {\it exact} random coding exponents, not just bounds. 
These relationships between the random coding exponents and the parameters
$\alpha$ and $\beta$ can, in principle, be
inverted (in a certain domain) in order to find the assignments of $\alpha$
and $\beta$ needed to satisfy given constraints on the exponents
of the FA and the MD probabilities.
For the sake of fairness, on the
other hand, it should also be made 
clear that since we consider only the random coding regime, these are merely
achievability results, with no converse bounds pertaining to optimal codes.

The outline of the paper is as follows.
In Section 2, we establish some notation conventions, provide some
preliminaries, and finally, formulate the problem.
In Section 3, we derive the optimum detector/decoder and discuss some of its
properties. In Section 4, we present our main theorem, which is about single--letter
formulas for the various error exponents. Finally, in Section 5, we prove this
theorem.

\section{Notation Conventions, Preliminaries and Problem Formulation}

\subsection{Notation Conventions and Preliminaries}

Throughout the paper, random variables will be denoted by capital
letters, specific values they may take will be denoted by the
corresponding lower case letters, and their alphabets, similarly as other sets,
will be denoted by calligraphic letters. Random
vectors and their realizations will be denoted,
respectively, by capital letters and the corresponding lower case letters,
both in the bold face font. Their alphabets will be superscripted by their
dimensions. For example, the random vector $\bX=(X_1,\ldots,X_n)$, ($n$ -- positive
integer) may take a specific vector value $\bx=(x_1,\ldots,x_n)$
in $\calX^n$, the $n$--th order Cartesian power of $\calX$, which is
the alphabet of each component of this vector.

For a given vector $\bx$, let $\hat{Q}_X$ denote\footnote{In our notation, 
we do not index $\hQ_X$ by $\bx$ because the underlying sequence $\bx$ will be
clear from the context.}
the empirical
distribution, that is, the vector $\{\hat{Q}_X(x),~x\in\calX\}$, where
$\hat{Q}_X(x)$ is the relative frequency of the letter $x$ in the vector
$\bx$. Let
$\calT_P$ denote the type class associated with $P$, that is, the set of
all sequences $\{\bx\}$ for which $\hQ_X=P$.
Similarly, for a pair of vectors
$(\bx,\by)$, the empirical joint distribution will be denoted by $\hat{Q}_{XY}$ or simply $\hQ$
for short. Conditional empirical distributions will be denoted by $\hQ_{X|Y}$
and $\hQ_{Y|X}$, the $y$--marginal by $\hQ_Y$, etc. Accordingly,
the empirical mutual information induced by $(\bx,\by)$ will be denoted by 
$I(\hQ_{XY})$ or $I(\hQ)$, 
the divergence between $\hQ_X$ and $P=\{P(x),~x\in\calX\}$ -- by
$\calD(\hQ_Y\|P)$, and
the conditional divergence between the empirical conditional
distribution $\hQ_{Y|X}$ and the channel $W=\{W(y|x)~x\in\calX,~y\in\calY\}$,
will be denoted by $\calD(\hQ_{Y|X}\|W|\hQ_X)$, that is,
\begin{equation}
\calD(\hQ_{Y|X}\|W|\hQ_X)=\sum_{x\in\calX}\hQ_X(x)\sum_{y\in\calY}\hQ_{Y|X}(y|x)\log
\frac{\hQ_{Y|X}(y|x)}{W(y|x)},
\end{equation}
and so on. The joint distribution induced by $\hQ_X$ and $\hQ_{Y|X}$ will be
denoted by $\hQ_X\times\hQ_{Y|X}$, and a similar notation will be used when the
roles of $X$ and $Y$ are switched.
The marginal of $X$, induced by $\hQ_Y$ and $\hQ_{X|Y}$ will be
denoted by $(\hQ_Y\times\hQ_{X|Y})_X$, and so on.
Similar notation conventions will apply, of course, to generic distributions
$Q_{XY}$, $Q_X$, $Q_Y$, $Q_{Y|X}$, and $Q_{X|Y}$, which are not necessarily
empirical distributions (without ``hats'').

The expectation operator will be denoted by $\bE\{\cdot\}$. Whenever there is room
for ambiguity, the underlying probability distribution will appears as a
subscript, e.g., $\bE_Q\{\cdot\}$.
Logarithms and exponents will be understood to be taken to the natural base
unless specified otherwise.
The indicator function will be denoted by $\calI(\cdot)$. 
Sets will normally be denoted by calligraphic letters. The complement of a set
$\calA$ will be denoted by $\overline{\calA}$.
The notation
$[t]_+$ will stand for $\max\{t,0\}$. For two positive sequences,
$\{a_n\}$ and $\{b_n\}$, the notation $a_n\exe b_n$ will mean asymptotic
equivalence in the exponential scale, that is,
$\lim_{n\to\infty}\frac{1}{n}\log(\frac{a_n}{b_n})=0$.
Similarly, $a_n\lexe b_n$ will mean
$\limsup_{n\to\infty}\frac{1}{n}\log(\frac{a_n}{b_n})\le 0$, and so on.
Throughout the sequel, we will make frequent use of the fact that
$\sum_{i=1}^{k_n} a_i(n) \exe \max_{1\le i\le k_n} a_i(n)$ as long as as
$\{a_i(n)\}$ are positive and $k_n\exe 1$. Accordingly, for $k_n$
sequences of positive random
variables $\{A_i(n)\}$, all defined on a common probability space, 
and a deterministic sequence $B_n$,
\begin{eqnarray}
\label{pullout}
\mbox{Pr}\left\{\sum_{i=1}^{k_n} A_i(n)\ge B_n\right\}
&\exe&\mbox{Pr}\left\{\max_{1\le i\le k_n}A_i(n)\ge B_n\right\}\nonumber\\
&=&\mbox{Pr}\bigcup_{i=1}^{k_n}\left\{A_i(n)\ge B_n\right\}\nonumber\\
&\exe&\sum_{i=1}^{k_n}\mbox{Pr}\left\{A_i(n)\ge B_n\right\}\nonumber\\
&\exe&\max_{1\le i\le k_n}\mbox{Pr}\left\{A_i(n)\ge B_n\right\},
\end{eqnarray}
provided that $B_n'\exe B_n$ implies $\mbox{Pr}\{A_i(n)\ge B_n'\}\exe
\mbox{Pr}\{A_i(n)\ge B_n\}$.\footnote{Consider the case where $B_n\exe e^{bn}$
($b$ being a constant independent of $n$)
and the exponent of $\mbox{Pr}\{A_i(n)\ge e^{bn}\}$ is a continuous function
of $b$.}
In simple words, summations and maximizations are equivalent and can be both ``pulled out outside''
$\mbox{Pr}\{\cdot\}$ without changing the exponential order, as long as
$k_n\exe 1$. By the same token,
\begin{eqnarray}
\label{intersect}
\mbox{Pr}\left\{\sum_{i=1}^{k_n} A_i(n)\le B_n\right\}&\exe&
\mbox{Pr}\left\{\max_{1\le i\le k_n} A_i(n)\le B_n\right\}\nonumber\\
&=&\mbox{Pr}\bigcap_{i=1}^{k_n} \{A_i(n)\le B_n\}.
\end{eqnarray}
Another fact that will be used extensively is that for a given set of
$M$ pairwise independent events $\{\calA_i\}_{i=1}^M$,
\begin{equation}
\label{shulman}
\mbox{Pr}\left\{\bigcup_{i=1}^M\calA_i\right\}\exe\min\left\{1,\sum_{i=1}^M
\mbox{Pr}\{\calA_i\}\right\}.
\end{equation}
The right--hand side (r.h.s.) is obviously the union bound, which holds true even if
the events are not pairwise independent. On the other hand, 
when multiplied by a factor of $1/2$, 
the r.h.s.\ becomes a lower bound to
$\mbox{Pr}\{\bigcup_{i=1}^M\calA_i\}$, provided that $\{A_i\}$ are pairwise
independent \cite[Lemma A.2]{Shulman03}, \cite[Lemma 1]{SBM07}.

\subsection{Problem Formulation}

Consider a discrete memoryless channel (DMC), characterized by a finite input alphabet
$\calX_0$, a finite out alphabet $\calY$ 
and a given matrix of single--letter transition
probabilities $\{W(y|x),~x\in\calX_0,~y\in\calY\}$. 
It is further assumed that $\calX_0$ contains a special symbol denoted by
`$0$', which
designates the channel input in the absence of transmission.
We shall denote $\calX=\calX_0\setminus\{0\}$ and
$Q_0(y)=W(y|x=0)$.

We assume an ensemble of random codes, where each codeword
is selected independently at random, uniformly within a type class
$\calT_P$. Let $\calC=\{\bx_1,\bx_2\ldots,\bx_M\}$,
$\bx_m\in\calX^n$, $m=1,\ldots,M$, $M=e^{nR}$ ($R$ being the
coding rate in nats per channel use), denote the (randomly chosen) code, which
is revealed to both the encoder and the decoder.

A detector/decoder, for a code operating in the setting of slotted asynchronous
communication, is a partition of $\calY^n$ into $M+1$ regions,
denoted $\calR_0,\calR_1,\ldots,\calR_M$. If $\by\in\calR_m$,
$m=1,2,\ldots,M$,
then the decoder decodes the message to be $m$.
If $\by\in\calR_0$, then the decoder declares that
nothing has been transmitted, that is, $\bx=0^n$ and then 
$\by$ is ``pure noise.''
The probability of decoding error (DE) is defined as
\begin{equation}
P_{\mbox{\tiny DE}}=\frac{1}{M}\sum_{m=1}^MW(\overline{\calR_m})=
\frac{1}{M}\sum_{m=1}^M\sum_{k\ne m}W(\calR_k|\bx_m),
\end{equation}
where the inner summation at the right--most side {\it includes} $k=0$.
The probability of false alarm (FA) is defined as
\begin{equation}
P_{\mbox{\tiny FA}}=Q_0(\overline{\calR_0})=\sum_{m=1}^MQ_0(\calR_m),
\end{equation}
and the probability of misdetection (MD) is defined as
\begin{equation}
P_{\mbox{\tiny MD}}=\frac{1}{M}\sum_{m=1}^M W(\calR_0|\bx_m).
\end{equation}
For a given code $\calC$, we are basically interested 
in achievable trade-offs between $P_{\mbox{\tiny
DE}}$, $P_{\mbox{\tiny FA}}$, and $P_{\mbox{\tiny MD}}$. Consider
the following problem:
\begin{eqnarray}
\label{min}
& &\mbox{minimize}~~~P_{\mbox{\tiny DE}}\nonumber\\
& &\mbox{subject to}~~P_{\mbox{\tiny FA}}\le \epsilon_{\mbox{\tiny FA}}\nonumber\\
& &~~~~~~~~~~~~~~~P_{\mbox{\tiny MD}}\le \epsilon_{\mbox{\tiny MD}}
\end{eqnarray}
where $\epsilon_{\mbox{\tiny FA}}$ and $\epsilon_{\mbox{\tiny MD}}$ are given
prescribed quantities, and it assumed that these two constraints are not
contradictory.\footnote{Note that there is some tension between 
$P_{\mbox{\tiny MD}}$ and $P_{\mbox{\tiny
FA}}$ as they are related via the Neyman--Pearson lemma. For a given
$\epsilon_{\mbox{\tiny FA}}$, the minimum achievable MD probability
is positive, in general. It is assumed then that the prescribed value of
$\epsilon_{\mbox{\tiny MD}}$ is not smaller than this minimum. In the problem
under consideration, it makes sense to relax the tension between the two
constraints to a certain extent, in order to allow some freedom to minimize $P_{DE}$ under these
constraints.}

Our goal is to find the optimum detector/decoder and then analyze the random
coding exponents associated with the resulting error probabilities.

\section{The Optimum Detector/Decoder}

Let us define the following detector/decoder:
\begin{eqnarray}
\calR_0^*&=&\left\{\by:~a\cdot\sum_{m=1}^M W(\by|\bx_m)+\max_m W(\by|\bx_m)\le
b\cdot Q_\star(\by)\right\}\\
\calR_m^*&=&\overline{\calR_0^*}\bigcap\left\{\by:~W(\by|\bx_m)> \max_{k\ne m}
W(\by|\bx_k)\right\},~~~~m=1,2,\ldots,M,
\end{eqnarray}
where ties are broken arbitrarily, and where $a\ge 0$ and $b\ge 0$ are
deterministic constants.
The following lemma establishes the optimality of the decision rule
$\calR^*=\{\calR_0^*,\calR_1^*,\ldots,\calR_M^*\}$ in the sense of the
trade-off among the probabilities $P_{\mbox{\tiny MD}}$, $P_{\mbox{\tiny
FA}}$ and $P_{\mbox{\tiny DE}}$. It tells us that there is no other
decision rule that simultaneously yields strictly smaller error probabilities
of all three kinds.
\begin{lemma}
Let $\calR^*=\{\calR_0^*,\calR_1^*,\ldots,\calR_M^*\}$ be as above and let
$\calR=\{\calR_0,\calR_1,\ldots,\calR_M\}$ be any another partition of $\calY^n$
into $M+1$ regions. If
\begin{equation}
Q_0(\overline{\calR_0})\le Q_0(\overline{\calR_0^*})
\end{equation}
and 
\begin{equation}
\frac{1}{M}\sum_{m=1}^MW(\calR_0|\bx_m)\le 
\frac{1}{M}\sum_{m=1}^MW(\calR_0^*|\bx_m),
\end{equation}
then
\begin{equation}
\frac{1}{M}\sum_{m=1}^MW(\overline{\calR_m^*}|\bx_m)\le 
\frac{1}{M}\sum_{m=1}^MW(\overline{\calR_m}|\bx_m).
\end{equation}
\end{lemma}

\noindent
{\it Proof.} We begin from the obvious observation 
that for a given choice of $\calR_0$, the optimum
choice of the other decision regions is always:
\begin{equation}
\label{optrm}
\calR_m=
\overline{\calR_0}\bigcap\left\{\by:~W(\by|\bx_m)> \max_{k\ne
m}W(\by|\bx_k)\right\},~~~~~~~m=1,2,\ldots,M.
\end{equation}
In other words, once a transmission has been detected, the best
decoding rule is the ML decoding rule. Similarly as in classical hypothesis
testing theory, this is true because the probability of correct decoding,
\begin{equation}
P_{\mbox{\tiny CD}}=\frac{1}{M}\sum_{m=1}^M\sum_{\by\in\calR_m}W(\by|\bx_m),
\end{equation}
is upper bounded by
\begin{equation}
P_{\mbox{\tiny CD}}\le\frac{1}{M}\sum_{m=1}^M\sum_{\by\in\calR_m}\max_kW(\by|\bx_k)
=\frac{1}{M}\sum_{\by\in\overline{\calR_0}}\max_mW(\by|\bx_m)
\end{equation}
and this bound is achieved by (\ref{optrm}). Thus, upon adopting
(\ref{optrm}) for a given choice of $\calR_0$, 
it remains to prove that the choice $\calR_0^*$ satisfies the assertion of the lemma.

The proof of this fact is similar to the proof of the Neyman--Pearson lemma.
Let $\calR_0^*$ be as above and let $\calR_0$ be another, competing rejection
region. First, observe that for every $\by\in\calY^n$
\begin{equation}
[\calI\{\by\in\calR_0^*\}-\calI\{\by\in\calR_0\}]\cdot\left[b\cdot
Q_0(\by)-a\cdot\sum_{m=1}^MW(\by|\bx_m)-\max_mW(\by|\bx_m)\right]\ge 0.
\end{equation}
This is true because, by definition of $\calR_0^*$, 
the two factors of the product at the left--hand side (l.h.s.) are
either both non--positive or both non--negative.
Thus, taking the summation over all $\by\in\calY^n$, we have:
\begin{eqnarray}
0&\le&\sum_{\by\in\calY^n}
[\calI\{\by\in\calR_0^*\}-\calI\{\by\in\calR_0\}]\cdot\left[b\cdot
Q_0(\by)-a\cdot\sum_{m=1}^MW(\by|\bx_m)-\max_mW(\by|\bx_m)\right]\nonumber\\
&=&b\cdot[Q_0(\calR_0^*)-Q_0(\calR_0)]-
a\cdot\left[\sum_{m=1}^MW(\calR_0^*|\bx_m)-\sum_{m=1}^MW(\calR_0|\bx_m)\right]-\nonumber\\
& &\left[\sum_{\by\in\calR_0^*}\max_mW(\by|\bx_m)-\sum_{\by\in\calR_0}\max_mW(\by|\bx_m)\right]
\end{eqnarray}
which yields
\begin{eqnarray}
& &\sum_{\by\in\calR_0^*}\max_mW(\by|\bx_m)-\sum_{\by\in\calR_0}\max_mW(\by|\bx_m)\nonumber\\
&\le&
b\cdot[Q_0(\calR_0^*)-Q_0(\calR_0)]-
a\cdot\left[\sum_{m=1}^MW(\calR_0^*|\bx_m)-\sum_{m=1}^MW(\calR_0|\bx_m)\right]\nonumber\\
&=&
b\cdot[Q_0(\overline{\calR_0})-Q_0(\overline{\calR_0^*})]+
a\cdot\left[\sum_{m=1}^MW(\calR_0|\bx_m)-\sum_{m=1}^MW(\calR_0^*|\bx_m)\right]
\end{eqnarray}
Since $a\ge 0$ and $b\ge 0$, it follows that 
\begin{equation}
Q_0(\overline{\calR_0})\le Q_0(\overline{\calR_0^*}) 
\end{equation}
and 
\begin{equation}
\frac{1}{M}\sum_{m=1}^MW(\calR_0|\bx_m)\le\frac{1}{M}\sum_{m=1}^MW(\calR_0^*|\bx_m)
\end{equation}
together imply that
\begin{equation}
\sum_{\by\in\calR_0^*}\max_mW(\by|\bx_m)\le
\sum_{\by\in\calR_0}\max_mW(\by|\bx_m)
\end{equation}
or equivalently,
\begin{equation}
\sum_{\by\in\overline{\calR_0^*}}\max_mW(\by|\bx_m)\ge
\sum_{\by\in\overline{\calR_0}}\max_mW(\by|\bx_m),
\end{equation}
which in turn yields
\begin{eqnarray}
\frac{1}{M}\sum_{m=1}^MW(\overline{\calR_m^*}|\bx_m)&\equiv&1-
\frac{1}{M}\sum_{\by\in\overline{\calR_0^*}}\max_mW(\by|\bx_m)\nonumber\\
&\le&
1-\frac{1}{M}\sum_{\by\in\overline{\calR_0}}\max_mW(\by|\bx_m)\nonumber\\
&\equiv&\frac{1}{M}\sum_{m=1}^MW(\overline{\calR_m}|\bx_m).
\end{eqnarray}
This completes the proof of Lemma 1. $\Box$

\noindent
{\bf Discussion.} At this point, two comments are in order.

\noindent
1. The results thus far hold for any given code $\calC$. As mentioned earlier, 
in this work, we analyze
the ensemble performance.
Specifically, let $\bar{P}_{\mbox{\tiny DE}}$, $\bar{P}_{\mbox{\tiny FA}}$, and
$\bar{P}_{\mbox{\tiny MD}}$ denote the corresponding ensemble averages of
$P_{\mbox{\tiny DE}}$, $P_{\mbox{\tiny FA}}$, and
$P_{\mbox{\tiny MD}}$, respectively. We will assess the random coding
exponents of these three probabilities. 
The constants $a$ and $b$ can be thought of as Lagrange multipliers that are
tuned to meet the given FA and MD constraints.
For these Lagrange multipliers
to have an impact on error exponents, we let them be exponential
functions of $n$, that is, $a=e^{n\alpha}$ and $b=e^{n\beta}$,
where $\alpha$
and $\beta$ are real numbers, independent of $n$. The rejection region is then
of the form
\begin{equation}
\label{finalR0}
\calR_0^*=\left\{\by:~e^{n\alpha}\sum_{m=1}^MW(\by|\bx_m)+\max_mW(\by|\bx_m)\le
e^{n\beta}Q_0(\by)\right\}.
\end{equation}
By the same token, we impose
exponential constraints on the FA and MD probabilities,
that is, $\epsilon_{\mbox{\tiny FA}}=e^{-nE_{\mbox{\tiny FA}}}$ and
$\epsilon_{\mbox{\tiny MD}}=e^{-nE_{\mbox{\tiny MD}}}$, where
$E_{\mbox{\tiny FA}} \ge 0$ and
$E_{\mbox{\tiny MD}} \ge 0$ are given numbers, independent of $n$.

\noindent
2. The detection/rejection rule defined by (\ref{finalR0}) involves a
linear combination of $\max_mW(\by|\bx_m)$ and $\sum_{m=1}^MW(\by|\bx_m)$, or
equivalently, the overall output
distribution induced by the code
\begin{equation}
Q_{\calC}(\by)\dfn\frac{1}{M}\sum_{m=1}^MW(\by|\bx_m).
\end{equation}
In this context, the intuition behind the 
optimality of this detection rule is not trivial (at least for the
author of this article), 
and as mentioned earlier, it is very different from
that of \cite{Wang10} and \cite{WCCW11}. It is instructive, nonetheless, to
examine some special cases.
The first observation is that for
$\alpha\ge 0$, the term $e^{n\alpha}\sum_{m=1}^MW(\by|\bx_m)$ dominates the
term $\max_mW(\by|\bx_m)$, and so, the rejection region is essentially
equivalent to 
\begin{equation}
\label{approxr0}
\calR_0'=\left\{\by:~e^{n\alpha}\sum_{m=1}^MW(\by|\bx_m)\le
e^{n\beta}Q_0(\by)\right\}=
\left\{\by:~e^{n(\alpha+R)}Q_{\calC}(\by)\le
e^{n\beta}Q_0(\by)\right\},
\end{equation}
which is exactly the Neyman--Pearson test between $Q_{\calC}(\by)$ and
$Q_0(\by)$. This means that $\alpha\ge 0$ corresponds to a regime of full
tension between the FA and the MD constraints (see footnote no.\ 2).
In this case, $E_{\mbox{\tiny FA}}$ and $E_{\mbox{\tiny MD}}$ are related via the
Neyman--Pearson lemma, and there are no degrees of freedom left for
minimizing $\bar{P}_{\mbox{\tiny DE}}$ (or equivalently, maximizing its
exponent). Indeed, the detection--rejection rule (\ref{approxr0}) depends
only on one degree of freedom, which is the difference $\alpha-\beta$, 
and hence so are the FA and MD error
exponents associated with it.
At the other extreme, where
$e^{n\alpha} \ll 1$, and the term $\max_mW(\by|\bx_m)$ dominates, the detection rule
becomes equivalent to 
\begin{equation}
\calR_0''=\left\{\by:~\max_mW(\by|\bx_m)\le
e^{n\beta}Q_0(\by)\right\}.
\end{equation}
In this case, the silent mode is essentially treated as corresponding to yet
another codeword -- $\bx_0=0^n$, although it still has a special stature due to the
factor $e^{n\beta}$. But for $\beta=0$, this ``silent codeword'' is just an additional codeword
with no special standing, and the decoding is completely ordinary. The interesting range is therefore
the range where $\alpha$ is negative, but not too small, where both 
$Q_{\calC}(\by)$ and $\max_mW(\by|\bx_m)$ play a considerable role.

\section{Performance}

In this section, we present our main theorem, which provides exact single--letter
characterizations for all three exponents as functions of $\alpha$ and
$\beta$. We first need some definitions. Let
\begin{equation}
d(x,y)\dfn\ln\left[\frac{Q_0(y)}{W(y|x)}\right],~~~~x\in\calX,~y\in\calY
\end{equation}
and denote $D(Q)=\bE_Qd(X,Y)$.
For a given output distribution $Q_Y=\{Q_Y(y),~y\in\calY\}$,
define\footnote{
Conceptually, $\bR(D,Q_Y)$ can be thought of as the rate--distortion function of
the ``source'' $P$ subject to a constrained reproduction distribution $Q_Y$ (or
vice versa), but note that the ``distortion measure'' $d(x,y)$ here is not necessarily
non--negative for all $(x,y)$.}
\begin{equation}
\bR(\Delta;Q_Y)\dfn\inf_{\{Q_{Y|X}:~D(Q)\le \Delta,~(P\times
Q_{Y|X})_Y=Q_Y\}}I(Q).
\end{equation}
Next, define
\begin{equation}
\mu(Q_Y,R)\dfn\min_{Q_{X|Y}\in\calQ_P,~I(Q)\le R}\{I(Q)+D(Q)\},
\end{equation}
\begin{equation}
\tilde{\bR}(\Delta,R;Q_Y)\dfn\left\{\begin{array}{ll}
\bR(\Delta;Q_Y)-R & \Delta\le \mu(Q_Y,R)-R\\
0 & \Delta> \mu(Q_Y,R)-R\end{array}\right.
\end{equation}
\begin{equation}
E_A\dfn\inf_{Q_Y}[\calD(Q_Y\|Q_0)+\tilde{\bR}(\alpha-\beta,R;Q_Y)],
\end{equation}
\begin{equation}
E_B\dfn\inf_{Q_Y}\left\{\calD(Q_Y\|Q_0)+[\bR(-\beta;Q_Y)-R]_+\right\},
\end{equation}
and
\begin{equation}
\label{efa}
E_{\mbox{\tiny
FA}}\dfn\min\{E_A,E_B\}.
\end{equation}
The inverse function of $\bR(D;Q_Y)$, will be denoted by $\bD(R;Q_Y)$, i.e.,
\begin{equation}
\bD(R;Q_Y)=\inf_{\{Q_{Y|X}:~I(Q)\le R,~(P\times
Q_{Y|X})_Y=Q_Y\}}D(Q).
\end{equation}
Also, let $R_1(Q_Y)$ and $D_1(Q_Y)$ denote $I(Q^*)$ and $D(Q^*)$,
where $Q^*$ minimizes $I(Q)+D(Q)$.
Now, let
\begin{equation}
\label{emd}
E_{\mbox{\tiny MD}}\dfn\inf\calD(Q_{Y|X}\|W|P)
\end{equation}
where the infimum is subject to the constraints:
\begin{enumerate}
\item $\bD(R;Q_Y)\le[\alpha]_+-\beta\le D(P\times Q_{Y|X})$
\item $D_1(Q_Y)\le[\alpha]_+-\beta$ implies $\bR([\alpha]_+-\beta;Q_Y)\ge
R-[-\alpha]_+$
\item $D_1(Q_Y)>[\alpha]_+-\beta$ implies $R_1(Q_Y)+D_1(Q_Y)\ge
R+\alpha-\beta$
\end{enumerate}
with $Q_Y=(P\times Q_{Y|X})_Y$.
Next define
\begin{equation}
E_1=\inf_{\{Q_{Y|X}:~D(P\times Q_{Y|X})\le[\alpha]_+-\beta\}}
\left\{\calD(Q_{Y|X}\|W|P)+\left[\bR(D(P\times Q_{Y|X});(P\times Q_{Y|X})_Y)-R\right]_+\right\},
\end{equation}
\begin{equation}
E_2=\inf_{Q_{Y|X}}
\left\{\calD(Q_{Y|X}\|W|P)+\left[\bR(\alpha-\beta;(P\times Q_{Y|X})_Y)-R\right]_+\right\},
\end{equation}
and finally,
\begin{equation}
\label{ede}
E_{\mbox{\tiny DE}}\dfn\min\{E_1,E_2,E_{\mbox{\tiny MD}}\}.
\end{equation}

\begin{theorem}
Let $W$ be a DMC and let $\calR^*$ be both defined as in Section 2.2.
Let the codewords of $\calC=\{\bx_1,\ldots,\bx_M\}$, $M=e^{nR}$, be selected independently at
random under the uniform distribution across a given type class $\calT_P$.
Then, the asymptotic exponents associated with $\bar{P}_{\mbox{\tiny FA}}$,
$\bar{P}_{\mbox{\tiny MD}}$, and
$\bar{P}_{\mbox{\tiny DE}}$ are given, respectively, by
$E_{\mbox{\tiny FA}}$,
$E_{\mbox{\tiny MD}}$, and
$E_{\mbox{\tiny DE}}$, as defined in eqs.\ (\ref{efa}), (\ref{emd}), and
(\ref{ede}).
\end{theorem}

\noindent
{\bf Discussion.}
As discussed in Section 3, we observe that for $\alpha\ge 0$, all three
exponents depend on $\alpha$ and $\beta$ only via the difference
$\alpha-\beta$. It is also seen that there is nothing to lose by replacing
a positive value of $\alpha$ by $\alpha=0$, as long as the difference
$\alpha-\beta$ is kept. For $\alpha < 0$, the various exponents depend on $\alpha$ and
$\beta$ individually, so there are two degrees of freedom to adjust both the
FA and the MD exponents to pre--specified values in a certain range.

It is instructive to find out the maximum achievable information rate for
which the average probability of decoding error still tends to zero, that is,
the smallest rate $R$ for which $E_{\mbox{\tiny DE}}=0$, for given
$E_{\mbox{\tiny MD}}$ and 
$E_{\mbox{\tiny FA}}$. This happens as soon as either $E_1=0$ or $E_2=0$.
The exponent $E_1$ vanishes for $R=\bR(D(P\times W);(P\times W)_Y)$.
But
\begin{eqnarray}
\bR(P\times W;(P\times W)_Y)&=&\min\{I(Q):~D(Q)\le D(P\times W),~(P\times
Q_{Y|X})_Y=(P\times W)_Y\}\nonumber\\
&\le& I(P\times W).
\end{eqnarray}
On the other hand, since $\calD(Q_{Y|X}\|W|P)\ge 0$, it is easy to see that
the constraint set $\{Q:~D(Q)\le D(P\times W),~(P\times Q_{Y|X})_Y=(P\times
W)_Y\}$ is a subset of $\{Q:~I(Q)\ge I(P\times W)\}$, and so,
\begin{equation}
\bR(P\times W;(P\times W)_Y)
\ge\min\{I(Q):~I(Q)\ge I(P\times W)\}=I(P\times W),
\end{equation}
therefore, $\bR(P\times W;(P\times W)_Y)=I(P\times W)$, which is the ordinary
achievable rate one would expect from a constant composition code of type
class $\calT_P$. The exponent $E_2$ vanishes at the rate
$\bR(\alpha-\beta;(P\times W)_Y)$
Therefore, there is no rate loss, compared to ordinary decoding, as long as 
\begin{equation}
\alpha-\beta
\le D(P\times W).
\end{equation}

\section{Proof of Theorem 1}

This section is divided into three subsections, each one devoted to the
analysis of one of the three error exponents.

\subsection{The False Alarm Error Exponent}

Let $\by$ be given and consider $\{\bX_m\}$ as random. Then,
\begin{eqnarray}
\bar{P}_{\mbox{\tiny FA}}(\by)&=&Q_0\left\{
e^{n\alpha}\cdot\sum_{m=1}^MW(\by|\bX_m)+\max_m W(\by|\bX_m)>
e^{n\beta}Q_0(\by)\right\}\\
&\exe&Q_0\left\{e^{n\alpha}\cdot\sum_{m=1}^MW(\by|\bX_m)
> e^{n\beta}Q_\star(\by)\right\}+ Q_0\left\{\max_m W(\by|\bX_m)> 
e^{n\beta}Q_\star(\by)\right\}\\
&=&Q_0\left\{\sum_{m=1}^MW(\by|\bX_m)
> e^{n(\beta-\alpha)}Q_0(\by)\right\}+ Q_0\left\{\max_m W(\by|\bX_m)> 
e^{n\beta}Q_\star(\by)\right\}\\
&\dfn& A(\by)+B(\by),
\end{eqnarray}
where we have used (\ref{pullout}).
It is sufficient now to show that $A=\bE\{A(\bY)\}\exe e^{-nE_A}$ and
$B=\bE\{B(\bY)\}\exe e^{-nE_B}$.
Now, for a given $\by$, let $N(\hQ|\by)$ be the number of codewords in
$\calC$ whose joint empirical distribution with $\by$ is
$\hat{Q}=\{\hat{Q}(x,y),~x\in\calX,~y\in\calY\}$.
Next, define 
\begin{equation}
f(\hQ)=\sum_{x,y}\hQ(x,y)\ln W(y|x) 
\end{equation}
and
\begin{equation}
g(\hQ_Y)=\sum_y\hQ_Y(y)\ln Q_\star(y)+\beta-\alpha.
\end{equation}
Then,
\begin{eqnarray}
A(\by)&=&Q_0\left\{\sum_{\hQ_{X|Y}}N(\hQ|\by)e^{nf(\hQ)}
> e^{ng(\hQ_Y)}\right\}\nonumber\\
&\exe&Q_0\left\{\max_{\hQ_{X|Y}}N(\hQ|\by)e^{nf(\hQ)}>
e^{ng(\hQ_Y)}\right\}\\
&=&Q_0\bigcup_{\hQ_{X|Y}}\left\{N(\hQ|\by)e^{nf(\hQ)}>
e^{ng(\hQ_Y)}\right\}\\
&\exe&\sum_{\hQ_{X|Y}}Q_0\left\{N(\hQ|\by)>
e^{n[g(\hQ_Y)-f(\hQ)}\right\}\\
&\exe&\max_{\hQ_{X|Y}}Q_0\left\{N(\hQ|\by)>
e^{nu(\hQ)}\right\},
\end{eqnarray}
where we have used again eq.\ (\ref{pullout}) and where we have defined
\begin{equation}
u(\hQ)\dfn
g(\hQ_Y)-f(\hQ)=\sum_{x,y\in\calX\times\calY}\hQ(x,y)\ln\frac{Q_0(y)}{W(y|x)}+\beta-\alpha
=D(\hQ)+\beta-\alpha.
\end{equation}
Now, since $N(\hQ|\by)$ is a Bernoulli
random variable pertaining to $e^{nR}$ trials and probability of success
of the exponential order of $e^{-nI(\hQ)}$, we have, similarly as in
\cite[Subsection 6.3]{Merhav09}
\begin{equation}
\mbox{Pr}\{N(\hQ|\by)\ge e^{nu(\hQ)}\}\exe
\exp\left\{-e^{n[u(\hQ)]_+}(n[I(\hQ)-R+[u(Q)]_+]-1)\right\},
\end{equation}
provided that for $u(\hQ)> 0$, $I(\hQ)-R+u(\hQ)> 0$ (otherwise,
$\mbox{Pr}\{N(\hQ|\by)\ge e^{nu(\hQ)}\}\to 1$).\footnote{
Note also that $\mbox{Pr}\{N(\hQ|\by)\ge e^{nu(\hQ)}\}=
\mbox{Pr}\{N(\hQ|\by)\ge e^{n[u(\hQ)]_+}\}$ since $N(\hQ|\by)$ is an integer
valued random variable.}
Therefore, the exponential rate $E(\hQ)$ of $\mbox{Pr}\{N(\hQ|\by)\ge
e^{nu(\hQ)}\}$ is as follows:
\begin{equation}
E(\hQ)=\left\{\begin{array}{ll}
[I(\hQ)-R]_+ & u(\hQ)\le 0\\
\infty & u(\hQ)> 0,~u(\hQ)> R-I(\hQ)\\
0 & u(\hQ)> 0,~u(\hQ)< R-I(\hQ)\end{array}\right.
\end{equation}
For a given $\hQ_Y$, let $\calQ_P$ be the set of $\{\hQ_{X|Y}\}$ such that
$(\hQ_Y\times\hQ_{X|Y})_X=P$. Then,
\begin{eqnarray}
\min_{\hQ_{X|Y}\in\calQ_P}E(\hQ)&=&\left\{\begin{array}{ll}
\infty & \forall \hQ_{X|Y}\in\calQ_P:~u(\hQ)> 0,~u(\hQ)>R-I(\hQ)\\
0 & \exists \hQ_{X|Y}\in\calQ_P:~0\le u(\hQ)\le R-I(\hQ)\\
0 & \exists \hQ_{X|Y}\in\calQ_P:~u(\hQ)\le 0,~I(\hQ)\le R\\
\min_{\{\hQ_{X|Y}\in\calQ_P:~u(\hQ)\le 0\}}[I(\hQ)-R]_+ &
\mbox{otherwise}\end{array}\right.\nonumber\\
&=&\left\{\begin{array}{ll}
\infty & \forall \hQ_{X|Y}\in\calQ_P:~u(\hQ)> [R-I(\hQ)]_+\\
0 & \exists \hQ_{X|Y}\in\calQ_P:~I(\hQ)\le \min\{R,R-u(\hQ)\}\\
\min_{\{\hQ_{X|Y}\in\calQ_P:~u(\hQ)\le 0\}}[I(\hQ)-R]_+ &
\mbox{otherwise}\end{array}\right.\nonumber
\end{eqnarray}
The condition for $\min_{\hQ_{X|Y}\in\calQ_P}E(\hQ)$ to vanish
becomes
\begin{eqnarray}
\alpha-\beta+R&\ge&\mu(\hQ_Y,R)=
\min_{\{\hQ_{X|Y}\in\calQ_P:~I(\hQ)\le R\}}[I(\hQ)+D(\hQ)]\nonumber\\
&=&\left\{\begin{array}{ll}
R+\bD(R;\hQ_Y) & R < R_1(\hQ_Y)\\
R_1(\hQ_Y)+D_1(\hQ_Y) & R \ge R_1(\hQ_Y)\end{array}\right.
\end{eqnarray}
The condition for an infinite exponent is as follows: For $u(\hQ)$ to be
non-negative for all $\hQ_{X|Y}$, we need
\begin{equation}
\alpha-\beta\le D_{\min}(\hQ_Y)\dfn 
\min_{\hQ_{X|Y}\in\calQ_P}D(\hQ).
\end{equation}
For $u(\hQ)\ge R-I(\hQ)$ for all $\hQ_{X|Y}\in\calQ_P$, we need
$\alpha-\beta+R<\mu(\hQ_Y,\infty)$.
Thus, in summary,
\begin{eqnarray}
\min_{\hQ_{X|Y}\in\calQ_P}E(\hQ)&=&\left\{\begin{array}{ll}
0 & \alpha-\beta \ge \mu(\hQ_Y,R)-R\\
\infty & \alpha-\beta < \min\{\mu(\hQ_Y,\infty)-R,D_{\min}(\hQ_Y)\}\\
\min_{\{\hQ_{X|Y}\in\calQ_P:~u(\hQ)\le 0\}}[I(\hQ)-R]_+ &
\mbox{elsewhere}\end{array}\right.\nonumber\\
&=&\left\{\begin{array}{ll}
0 & \alpha-\beta \ge \mu(\hQ_Y,R)-R\\
\infty & \alpha-\beta < \min\{\mu(\hQ_Y,\infty)-R,D_{\min}(\hQ_Y)\}\\
\left[\bR(\alpha-\beta;\hQ_Y)-R\right]_+ &
\mbox{elsewhere}\end{array}\right.\nonumber\\
&=&\left\{\begin{array}{ll}
\bR(\alpha-\beta;\hQ_Y)-R & \alpha-\beta< \mu(\hQ_Y,R)-R\\
0 & \alpha-\beta\ge \mu(\hQ_Y,R)-R\end{array}\right.\nonumber\\
&=&\tilde{\bR}(\alpha-\beta,R;\hQ_Y),
\end{eqnarray}
where we have used the convention that the minimum over an empty set is
infinity and the fact that $\bD(R;\hQ_Y)\ge \mu(Q_Y,R)-R$.
For the overall exponent associated with $A$, we need to average over $\bY$,
which gives $A\exe e^{-nE_A}$ with
\begin{equation}
E_A=\min_{Q_Y}\{\calD(Q_Y\|Q_0)+\tilde{\bR}(\alpha-\beta,R;Q_Y)\}.
\end{equation}

Moving on to the analysis of $B(\by)$,
\begin{eqnarray}
B(\by)&=&Q_0\left\{\max_m W(\by|\bX_m)> e^{n\beta}Q_0(\by)\right\}\\
&=&Q_0\bigcup_{m=1}^M\left\{W(\by|\bX_m)> e^{n\beta}Q_0(\by)\right\}\\
&\exe&\min\left\{1,M\cdot Q_0\{W(\by|\bX_1)> e^{n\beta}Q_0(\by)\}\right\},
\end{eqnarray}
where in the last line, we have used (\ref{shulman}).
Now,
\begin{equation}
Q_0\{W(\by|\bX_1)> e^{n\beta}Q_0(\by)\}\exe e^{-nI_0(\hQ_Y)},
\end{equation}
where
\begin{eqnarray}
I_0(\hQ_Y)&=&\min_{\hQ_{X|Y}}\left\{I(\hQ):~D(\hQ)\le -\beta,
~~\hQ_{X|Y}\in\calQ_P
\right\}\nonumber\\
&=&\bR(-\beta;\hQ_Y).
\end{eqnarray}
Thus, $B\exe e^{-nE_B}$ with
\begin{equation}
E_B=\min_{Q_Y}\{\calD(Q_Y\|Q_0)+[\bR(-\beta;Q_Y)-R]_+\}.
\end{equation}

\subsection{The Misdetection Error Exponent}

Without loss of generality, we will assume that $\bX_1=\bx_1$ was transmitted.
We first condition on $\bx_1$ and $\by$.
\begin{eqnarray}
\bar{P}_{\mbox{\tiny MD}}(\bx_1,\by)&=&
\mbox{Pr}\left\{e^{n\alpha}\sum_{m=1}^MW(\by|\bX_m)+\max_mW(\by|\bX_m)\le e^{n\beta}Q_\star(\by)
\bigg|\bX_1=\bx_1,\by\right\}\nonumber\\
&=&\mbox{Pr}\left\{e^{n\alpha}\sum_{m=1}^MW(\by|\bX_m)+\right.\nonumber\\
& &\left.\max\{W(\by|\bx_1),\max_{m>1}W(\by|\bX_m)\}\le e^{n\beta}Q_\star(\by)
\bigg|\bX_1=\bx_1,\by\right\}\nonumber\\
&\exe&\mbox{Pr}\left\{e^{n\alpha}\left[W(\by|\bx_1)+\sum_{m>1}W(\by|\bX_m)\right]+W(\by|\bx_1)+
\max_{m>1}W(\by|\bX_m)\le e^{n\beta}Q_\star(\by)
\bigg|\bx_1,\by\right\}\nonumber\\
&\exe&\mbox{Pr}\left\{e^{n[\alpha]_+}W(\by|\bx_1)+e^{n\alpha}\sum_{m>1}W(\by|\bX_m)+
\max_{m>1}W(\by|\bX_m)\le e^{n\beta}Q_\star(\by)
\bigg|\bx_1,\by\right\}\nonumber\\
&\exe&\mbox{Pr}\left\{e^{n[\alpha]_+}W(\by|\bx_1)< e^{n\beta}Q_\star(\by),
e^{n\alpha}\sum_{m>1}W(\by|\bX_m)+
\max_{m>1}W(\by|\bX_m)\le e^{n\beta}Q_\star(\by)
\bigg|\bx_1,\by\right\}\nonumber\\
&=&\calI\left\{e^{n[\alpha]_+}W(\by|\bx_1)< e^{n\beta}Q_\star(\by)\right\}\times\nonumber\\
& & \mbox{Pr}\left\{e^{n\alpha}\sum_{m>1}W(\by|\bX_m)+
\max_{m>1}W(\by|\bX_m)\le e^{n\beta}Q_\star(\by)
\bigg|\bx_1,\by\right\}\nonumber\\
&\dfn&C\cdot D.
\end{eqnarray}
Using the identity 
\begin{equation}
\max_{m>1}W(\by|\bx_m)\equiv\max_{\hQ_{X|Y}}\calI\{N(\hQ|\by)\ge 1\}\cdot
e^{nf(\hQ)} 
\end{equation}
(where now $N(\hQ|\by)$ does not count $\bx_1$), we now have 
\begin{eqnarray}
D&=&\mbox{Pr}\left\{e^{n\alpha}\sum_{m>1}W(\by|\bX_m)+
\max_{m>1}W(\by|\bX_m)\le e^{n\beta}Q_0(\by)
\bigg|\bx_1,\by\right\}\\
&=&\mbox{Pr}\left\{e^{n\alpha}\sum_{\hQ_{X|Y}}N(\hQ|\by)e^{nf(\hQ)}+
\max_{\hQ_{X|Y}}\calI\{N(\hQ|\by)\ge 1\}\cdot e^{nf(\hQ)}\le
e^{n[g(\hQ_Y)+\alpha]}
\bigg|\bx_1,\by\right\}\\
&\exe&\mbox{Pr}\left\{e^{n\alpha}\sum_{\hQ_{X|Y}}N(\hQ|\by)e^{nf(\hQ)}+
\sum_{\hQ_{X|Y}}\calI\{N(\hQ|\by)\ge 1\}e^{nf(\hQ)}\le e^{n[g(\hQ_Y)+\alpha]}
\bigg|\bx_1,\by\right\}\\
&=&\mbox{Pr}\left\{\sum_{\hQ_{X|Y}}[e^{n\alpha}N(\hQ|\by)+
\calI\{N(\hQ|\by)\ge 1\}]e^{nf(\hQ)}\le e^{n[g(\hQ_Y)+\alpha]}
\bigg|\bx_1,\by\right\}\\
&\exe&\mbox{Pr}\left\{\max_{\hQ_{X|Y}}[e^{n\alpha}N(\hQ|\by)+
\calI\{N(\hQ|\by)\ge 1\}]e^{nf(\hQ)}\le e^{n[g(\hQ_Y)+\alpha]}
\bigg|\bx_1,\by\right\}\\
&=&\mbox{Pr}\bigcap_{\hQ_{X|Y}}\left\{e^{n\alpha}N(\hQ|\by)+
\calI\{N(\hQ|\by)\ge 1\}\le e^{n[u(\hQ)+\alpha]}
\bigg|\bx_1,\by\right\}\\
&=&\mbox{Pr}\bigcap_{\hQ_{X|Y}}\left\{N(\hQ|\by)\le
e^{nv(\hQ)}
\bigg|\bx_1,\by\right\},
\end{eqnarray}
where
\begin{equation}
v(\hQ)=\left\{\begin{array}{ll}
u(\hQ) & u(\hQ)+\alpha > 0\\
-\infty & u(\hQ)+\alpha \le 0\end{array}\right.
\end{equation}
Now, if there exists at least one $\hQ_{X|Y}\in\calQ_P$ for which $I(\hQ) < R$ and
$R-I(\hQ)>v(\hQ)$, then this $\hQ_{X|Y}$ alone is responsible for a
double exponential decay of $D$ (because then the event in question would be
a large deviations event whose probability decays exponentially with
$M=e^{nR}$, thus double--exponentially with $n$), let alone the intersection over all
$\{\hQ_{X|Y}\}$. The condition for this to happen is
$R>R_0(\hQ_Y)\dfn\min_{Q_{X|Y}\in\calQ_P}\max\{I(\hQ),I(\hQ)+v(\hQ)\}$.
Conversely, if for every $\hQ$ with $\hQ_{X|Y}\in\calQ_P$, we have $I(\hQ) > R$ or
$R-I(\hQ)<v(\hQ)$, that is, $R<R_0(\hQ_Y)$, then $D$ is close to 1 since the intersection is
over a sub--exponential number of events with very high probability. It follows that $D$ behaves like
$\calI\{R_0(\hQ_Y)>R\}$, Thus, 
\begin{eqnarray}
P_{\mbox{\tiny MD}}&\exe&\bE\calI\left\{R_0(\hQ_Y)>R,~
W(\bY|\bX_1)\le
e^{n(\beta-[\alpha]_+)}Q_0(\bY)\right\}\nonumber\\
&=&\exp\left[-n\inf_{Q_{Y|X}\in\calQ_P}\left\{\calD(Q_{Y|X}\|W|P):~R_0(Q_Y)>R,~
D(Q)>[\alpha]_+-\beta\right\}\right].
\end{eqnarray}
Now, let us take a closer look at $R_0(Q_Y)$:
\begin{eqnarray}
\max\{I(Q),I(Q)+v(Q)\}&=&\left\{\begin{array}{ll}
\max\{I(Q),I(Q)+u(Q)\} & u(Q)>-\alpha\\
I(Q) & u(Q)\le-\alpha\end{array}\right.\\
&=&I(Q)+u(Q)\cdot\calI\{u(Q)>[-\alpha]_+\}.
\end{eqnarray}
Thus,
\begin{eqnarray}
R_0(Q)&=&\min_{Q_{X|Y}\in\calQ_P}[I(Q)+u(Q)\cdot\calI\{u(Q)>[-\alpha]_+\}]\\
&=&\min\left\{\min_{Q_{X|Y}\in\calQ_P:~u(Q)\le[-\alpha]_+}I(Q),
\min_{Q_{X|Y}\in\calQ_P:~u(Q)>[-\alpha]_+}[I(Q)+u(Q)]\right\}.
\end{eqnarray}
Now,
\begin{eqnarray}
\min_{Q_{X|Y}\in\calQ_P:~u(Q)\le[-\alpha]_+}I(Q)&=&\bR(\alpha+[-\alpha]_+-\beta;Q_Y)\\
&=&\bR([\alpha]_+-\beta;Q_Y)
\end{eqnarray}
and
\begin{eqnarray}
& &\min_{Q_{X|Y}\in\calQ_P:~u(Q)>[-\alpha]_+}[I(Q)+u(Q)]\\
&=&\beta-\alpha+\min_{Q_{X|Y}\in\calQ_P:~D(Q)>[\alpha]_+-\beta}[I(Q)+D(Q)]\\
&=&\beta-\alpha+\left\{\begin{array}{ll}
R_1(Q_Y)+D_1(Q_Y) & [\alpha]_+-\beta < D_1(Q_Y)\\
\bR([\alpha]_+-\beta;Q_Y)+[\alpha]_+-\beta & \mbox{otherwise}
\end{array}\right.\\
&=&\left\{\begin{array}{ll}
R_1(Q_Y)+D_1(Q_Y) +\beta-\alpha & [\alpha]_+-\beta < D_1(Q_Y)\\
\bR([\alpha]_+-\beta;Q_Y)+[\alpha]_+-\alpha & \mbox{otherwise}
\end{array}\right.\\
&=&\left\{\begin{array}{ll}
R_1(Q_Y)+D_1(Q_Y) +\beta-\alpha & [\alpha]_+-\beta < D_1(Q_Y)\\
\bR([\alpha]_+-\beta;Q_Y)+[-\alpha]_+ & \mbox{otherwise}
\end{array}\right.
\end{eqnarray}
Thus,
\begin{equation}
E_{\mbox{\tiny MD}}=\inf\calD(Q_{Y|X}\|W|P), 
\end{equation}
where the infimum is over all $\{Q_{Y|X}\}$ that 
satisfies the following conditions:
\begin{enumerate}
\item $\bD(R;Q_Y)\le[\alpha]_+-\beta\le D(P\times Q_{Y|X})$
\item $D_1(Q_Y)\le[\alpha]_+-\beta$ implies $\bR([\alpha]_+-\beta;Q_Y)\ge
R-[-\alpha]_+$
\item $D_1(Q_Y)>[\alpha]_+-\beta$ implies $R_1(Q_Y)+D_1(Q_Y)\ge
R+\alpha-\beta$
\end{enumerate}
where $Q_Y=(P\times Q_{Y|X})_Y$.

\subsection{The Decoding Error Exponent}

Let us denote
\begin{equation}
\Omega_m\dfn\left\{\by:~W(\by|\bx_m)>\max_{k\ne m}W(\by|\bx_k)\right\}.
\end{equation}
Then, for $m\ge 1$, $\calR_m^*=\overline{\calR_0^*}\cap\Omega_m$.
For a given code, the probability of decoding error is given by
\begin{eqnarray}
P_{\mbox{\tiny DE}}&=&\frac{1}{M}\sum_{m=1}^MW(\overline{\calR_m^*}|\bx_m)\\
&=&\frac{1}{M}\sum_{m=1}^MW(\calR_0^*\cup\overline{\Omega_m}|\bx_m)\\
&=&\frac{1}{M}\sum_{m=1}^MW(\overline{\calR_0^*}\cap\overline{\Omega_m}|\bx_m)+
\frac{1}{M}\sum_{m=1}^MW(\calR_0^*|\bx_m).
\end{eqnarray}
Upon taking the ensemble average, the second term becomes
$\bar{P}_{\mbox{\tiny MD}}$, which we have already analyzed in the previous
subsection. Its error exponent, $E_{\mbox{\tiny MD}}$, indeed appears as one of the
arguments of the $\min\{\cdot\}$ operator in eq.\ (\ref{ede}), and so, it remains
to show that the exponent of the ensemble average of the first term is $\min\{E_1,E_2\}$.
Let $\bX_1=\bx_1$ be transmitted and let $\bY=\by$ be received.
As before, we first condition on $(\bx_1,\by)$.
\begin{eqnarray}
\mbox{Pr}\{\overline{\calR_0^*}\cap\overline{\Omega_1}|\bx_1,\by\}
&=&\mbox{Pr}\left\{e^{n\alpha}\sum_m
W(\by|\bX_m)+\max_mW(\by|\bX_m)>e^{n\beta}Q_\star(\by),\right.\nonumber\\
& &\left.~\max_{m>1}W(\by|\bX_m)\ge
W(\by|\bx_1)\bigg|\bx_1,\by\right\}\nonumber\\
&\exe&\mbox{Pr}\left\{e^{n[\alpha]_+}W(\by|\bx_1)+e^{n\alpha}\sum_{m>1}
W(\by|\bX_m)+\max_{m>1}W(\by|\bX_m)>e^{n\beta}Q_\star(\by),\right.\nonumber\\
& &\left.~\max_{m>1}W(\by|\bX_m)\ge
W(\by|\bx_1)\bigg|\bx_1,\by\right\}\nonumber\\
&\exe&A(\bx_1,\by)+B(\bx_1,\by)+C(\bx_1,\by)
\end{eqnarray}
where
\begin{equation}
A(\bx_1,\by)=\calI\left\{W(\by|\bx_1)\ge e^{n(\beta-[\alpha]_+)}Q_\star(\by)\right\}
\cdot\mbox{Pr}\left\{
\max_{m>1}W(\by|\bX_m)\ge
W(\by|\bx_1)\bigg|\bx_1,\by\right\},
\end{equation}
\begin{equation}
B(\bx_1,\by)=\mbox{Pr}\left\{\sum_{m>1}W(\by|\bX_m)\ge
e^{n(\beta-\alpha)}Q_\star(\by),~\max_{m>1}W(\by|\bX_m)\ge
W(\by|\bx_1)\bigg|\bx_1,\by\right\},
\end{equation}
and
\begin{equation}
C(\bx_1,\by)=\mbox{Pr}\left\{\max_{m>1}W(\by|\bX_m)\ge
\max\{e^{n\beta}Q_\star(\by),~W(\by|\bx_1)\}\bigg|\bx_1,\by\right\}.
\end{equation}
We next analyze each one of these terms.
First, observe that for a given constant $S$ (which may depend on the given
$\bx_1$ and $\by$), we have
\begin{eqnarray}
\mbox{Pr}\left\{
\max_{m>1}\frac{W(\by|\bX_m)}{Q_\star(\by)}\ge
e^{-nS}\bigg|\bx_1,\by\right\}
&\exe&\min\left\{1, e^{nR}\cdot\mbox{Pr}\left\{\frac{W(\by|\bX_2)}{Q_\star(\by)}>
e^{-nS}|\bx_1,\by\right\}\right\}\\
&\exe&\min\left\{1, e^{nR}\cdot\mbox{Pr}\left\{\frac{W(\by|\bX_2)}{Q_\star(\by)}>
e^{-nS}|\bx_1,\by\right\}\right\}\\
&\exe& \exp\{-n[\bR(S,\hQ_Y)-R]_+\}.
\end{eqnarray}
In our case, $S=D(\tQ)$, where $\tQ$ is the empirical
joint distribution of $\bx_1$ and $\by$. Thus,
\begin{eqnarray}
A&\dfn& \bE\{A(\bX_1,\bY)\}\nonumber\\
&\exe&\exp\left[-n\min_{\{Q:~Q_{X|Y}\in\calQ_P:~D(Q)\le
[\alpha]_+-\beta\}}\{\calD(Q_{Y|X}\|W|P)+[\bR(D(Q),Q_Y)-R]_+\}\right]\nonumber\\
&=& e^{-nE_1}.
\end{eqnarray}
Concerning $C(\bx_1,\by)$, we similarly have:
\begin{eqnarray}
C(\bx_1,\by)&=&\mbox{Pr}\left\{\max_{m>1}\frac{W(\by|\bX_m)}{Q_\star(\by)}\ge
\max\left\{e^{n\beta},~\frac{W(\by|\bx_1)}{Q_\star(\by)}\right\}\bigg|\bx_1,\by\right\}\nonumber\\
&\exe&\min\left\{1,e^{nR}\cdot\mbox{Pr}\left\{\frac{W(\by|\bX_2)}{Q_\star(\by)}\ge
\max\left\{e^{n\beta},~\frac{W(\by|\bx_1)}
{Q_\star(\by)}\right\}\bigg|\bx_1,\by\right\}\right\}\nonumber\\
&\exe& \exp\left\{-n[\bR(\min\{-\beta,D(\tQ)\};\hQ_y)-R]_+\right\},
\end{eqnarray}
and so,
\begin{eqnarray}
C&\dfn& \bE\{C(\bX_1,\bY)\}\nonumber\\
&\exe&
\exp\left\{-n\min_{Q:~Q_{X|Y}\in\calQ_P}\{\calD(Q_{Y|X}\|W|P)+
[\bR(\min\{-\beta,D(Q)\};Q_Y)-R]_+\}\right\}\nonumber\\
&\lexe& e^{-nE_1},
\end{eqnarray}
therefore, $C$ is always dominated by $A$.
It remains then to show that $B=\bE\{B(\bX_1,\bY)\}\exe e^{-nE_2}$.
First, for given $(\bx_1,\by)$,
\begin{eqnarray}
B(\bx_1,\by)&\exe&\mbox{Pr}\left\{\sum_{\hQ_{X|Y}}N(\hQ|\by)e^{nf(\hQ)}\ge e^{ng(\hQ)},~
\sum_{\hQ_{X|Y}}\calI\{N(\hQ|\by)\ge 1\}\cdot e^{nf(\hQ)}\ge
e^{nf(\tQ)}\right\}\nonumber\\
&\exe&\mbox{Pr}\left\{\max_{\hQ_{X|Y}}N(\hQ|\by)e^{nf(\hQ)}\ge e^{ng(\hQ)},~
\max_{\hQ_{X|Y}}\calI\{N(\hQ|\by)\ge 1\}\cdot e^{nf(\hQ)}\ge
e^{nf(\tQ)}\right\}\nonumber\\
&\exe&\mbox{Pr}\left[\bigcup_{\hQ_{X|Y}}\{N(\hQ|\by)\ge
e^{nu(\hQ)}\}\right]\bigcap\left[\bigcup_{\hQ_{X|Y}}\{\calI\{N(\hQ|\by)\ge 1\}\ge
e^{n[f(\tQ)-f(\hQ)]}\}\right]\nonumber\\
&=&\mbox{Pr}\left[\bigcup_{\hQ_{X|Y}}\{N(\hQ|\by)\ge
e^{nu(\hQ)}\}\right]\bigcap\left[\bigcup_{\hQ_{X|Y}:~f(\tQ)\le f(\hQ)}\{\calI\{N(\hQ|\by)\ge 1\}\ge
e^{n[f(\tQ)-f(\hQ)]}\}\right]\nonumber\\
&=&\mbox{Pr}\bigcup_{\{\hQ_{X|Y},Q_{X|Y}':~f(\tQ)\le f(Q')\}}\left\{N(\hQ|\by)\ge
e^{nu(\hQ)},~N(Q'|\by)\ge 1\right\}\nonumber\\
&\exe&\mbox{Pr}\bigcup_{\hQ}\left\{N(\hQ|\by)\ge
e^{n[u(\hQ)]_+}\right\}+\nonumber\\
& &\sum_{\hQ_{X|Y}\ne Q_{X|Y}':~f(\tQ)\le f(Q')}\mbox{Pr}\{N(\hQ|\by)\ge 
e^{n[u(\hQ)]_+},~N(Q'|\by)\ge 1\}\nonumber\\
&\exe&\max_{\hQ_{X|Y}}\mbox{Pr}\left\{N(\hQ|\by)\ge
e^{n[u(\hQ)]_+}\right\}+\nonumber\\
& &\max_{\hQ_{X|Y}\ne Q_{X|Y}':f(\tQ)\le f(Q')}\mbox{Pr}\{N(\hQ|\by)\ge
e^{n[u(\hQ)]_+},~N(Q'|\by)\ge
1\}\nonumber\\
&\exe&\max_{\hQ_{X|Y}}\mbox{Pr}\{N(\hQ|\by)\ge
e^{n[u(\hQ)]_+}\}\nonumber\\
&=&\exp\{-n[\bR(\alpha-\beta;\hQ_Y)-R]_+\}.
\end{eqnarray}
where the last passage follows from an analysis almost identical to that of
$E_A$ in Subsection 5.1.
Thus,
\begin{equation}
B=\bE\{B(\bX_1,\bY)\}=\exp\{-n\min_{Q_{Y|X}}\{\calD(Q_{Y|X}\|W|P)+
[\bR(\alpha-\beta;Q_Y)-R]_+\}=e^{-nE_2}.
\end{equation}


\end{document}